\documentclass[12pt, reqno]{amsart}

\usepackage[margin=.9in, a4paper, foot=.3in]{geometry}

\let\oldtocsection=\tocsection

\let\oldtocsubsection=\tocsubsection

\let\oldtocsubsubsection=\tocsubsubsection

\renewcommand{\tocsection}[2]{\hspace{0em}\oldtocsection{#1}{#2}}
\renewcommand{\tocsubsection}[2]{\hspace{1em}\oldtocsubsection{#1}{#2}}
\renewcommand{\tocsubsubsection}[2]{\hspace{2em}\oldtocsubsubsection{#1}{#2}}

\makeatletter
\def\section{\@startsection{section}{1}%
	\z@{-.7\linespacing\@plus -\linespacing}{.5\linespacing}%
	{\normalfont\scshape\centering}}
\def\subsection{\@startsection{subsection}{2}%
	\z@{-.5\linespacing\@plus -.7\linespacing}{.5em}%
	{\normalfont\bfseries\mathversion{bold}}}
\makeatother

\usepackage{cite}

\usepackage{graphicx, color}

\definecolor{darkblue}{rgb}{0,0,.6}
\definecolor{darkred}{rgb}{.6,0,0}
\definecolor{darkgreen}{rgb}{0,0.6,0}

\usepackage{hyperref}
\hypersetup {
    pdfstartview=FitH,
    colorlinks,
    citecolor=darkgreen,
    linkcolor=darkred,
    urlcolor=darkblue
}

\usepackage[noBBpl,slantedGreek]{mathpazo}
\usepackage[mathcal]{euscript}

\allowdisplaybreaks

\numberwithin{equation}{section}

\newcommand {\ms}{{\mathstrut}} 

\newcommand {\bbC}{\mathbb C}
\newcommand {\bbN}{\mathbb N}
\newcommand {\bbZ}{\mathbb Z}

\newcommand {\calC}{\mathcal C}

\newcommand {\gothh}{\mathfrak h}
\newcommand {\gothk}{\mathfrak k}

\usepackage{bm}

\newcommand {\bmm}{\bm m}

\newcommand {\hlpo}{\mathfrak h_{l + 1}}
\newcommand {\klpo}{\mathfrak k_{l + 1}}
\newcommand {\lsllpo}{\mathcal L(\mathfrak{sl}_{l + 1})}

\newcommand {\Osc}{\mathrm{Osc}}
\newcommand {\slii}{\mathfrak{sl}_2}
\newcommand {\sliii}{\mathfrak{sl}_3}\newcommand {\gllpo}{\mathfrak{gl}_{l + 1}}
\newcommand {\sllpo}{\mathfrak{sl}_{l + 1}}
\newcommand {\thlpo}{\widetilde{\mathfrak h}_{l + 1}}
\newcommand {\tlsllpo}{\widetilde{\mathcal L}(\mathfrak{sl}_{l + 1})}
\newcommand {\uqbp}{\mathrm U_q(\mathfrak b_+)}
\newcommand {\uqbm}{\mathrm U_q(\mathfrak b_-)}
\newcommand {\uqgllpo}{\mathrm U_q(\mathfrak{gl}_{l + 1})}
\newcommand {\uqsllpo}{\mathrm U_q(\mathfrak{sl}_{l + 1})}
\newcommand {\uqlslii}{\mathrm U_q(\mathcal L(\mathfrak{sl}_2))}
\newcommand {\uqlsliii}{\mathrm U_q(\mathcal L(\mathfrak{sl}_3))}
\newcommand {\uqlsllpo}{\mathrm U_q(\mathcal L(\mathfrak{sl}_{l + 1}))}
\newcommand {\uqtlsllpo}{\mathrm U_q(\widetilde{\mathcal L}(\mathfrak{sl}_{l + 1}))}

\title[Quantum groups, Verma modules and $q$-oscillators: General linear case]{Quantum groups, Verma modules and $q$-oscillators: \\ General linear case}

\author[Kh. S. Nirov]{\vskip .2em Khazret S. Nirov}
\address{Institute for Nuclear Research of the Russian Academy of Sciences, 60th October Ave 7a, 117312 Moscow, Russia}
\curraddr{Mathematics and Natural Sciences, University of Wuppertal, 42097 Wuppertal, Germany}
\email{nirov@uni-wuppertal.de}

\author[A. V. Razumov]{Alexander V. Razumov}
\address{Institute for High Energy Physics, NRC "Kurchatov Institute", 142281 Protvino, Mos\-cow region, Russia}
\email{Alexander.Razumov@ihep.ru}

\begin{document}

\addtolength {\jot}{3pt}

\begin{abstract}
The Verma modules over the quantum groups $\mathrm U_q(\mathfrak{gl}_{l + 1})$ for arbitrary values of $l$ are analysed. The explicit expressions for the action of the generators on the elements of the natural basis are obtained. The corresponding representations of the quantum loop algebras $\mathrm U_q(\mathcal L(\mathfrak{sl}_{l + 1}))$ are constructed via Jimbo's homomorphism. This allows us to find certain representations of the positive Borel subalgebras of $\mathrm U_q(\mathcal L(\mathfrak{sl}_{l + 1}))$ as degenerations of the shifted representations. The latter are the representations used in the construction of the 
so-called $Q$-operators in the theory of quantum integrable systems. The interpretation 
of the corresponding simple quotient modules in terms of representations of the $q$-deformed oscillator algebra is given.
\end{abstract}

\maketitle

\tableofcontents

\section{Introduction} \label{s:1}

One of the most advanced methods of investigation of quantum integrable systems was developed on the basis of quantum groups \cite{Dri85, Dri87, Jim85}. The latter are, in a sense, the objects which have replaced the classical Lie groups and Lie algebras within the framework of the group-theoretic, or algebraic, approach to physical models. The primary problem in the study of quantum integrable systems is to describe the spectrum of the corresponding transfer matrices. This task reduces to the examination of functional relations in the system of transfer and $Q$-operators, being a substitution of the Bethe ansatz equations \cite{Bax72a}.\footnote{Actually, there is a vast list of interesting papers on this subject, see here some part of it \cite{KluPea92, KunNakSuz94, Koj08, BazLukMenSta10, Tsu10, FraLukMenSta11, KazLeuTsu12, AleKazLeuTsuZab13, Tsu13a, FraLukMenSta13}.}

In the quantum-group formalism the derivation of the functional relations is based on a thorough analysis of the appropriate representations of quantum groups and their Borel subalgebras. For the case of the quantum groups related to the Lie algebras $\slii$ and $\sliii$ the corresponding work has been carried out in the papers \cite{BazLukZam96, BazLukZam97, BazLukZam99, BazHibKho02, BooGoeKluNirRaz13, BooGoeKluNirRaz14a, NirRaz14, BooGoeKluNirRaz14b}.

In this paper we consider the case of the quantum groups $\uqlsllpo$, $l = 1, 2, \ldots,$ see section~\ref{s:5} for the definition. These quantum groups are deformations of enveloping algebras of the loop algebras of Lie algebras $\lsllpo$. Presently, in this case it is common instead of the term a quantum group to use the term a quantum loop algebra. From the point of view of quantum integrable systems the most interesting representations of $\uqlsllpo$ are those which are obtained from the Verma modules over the quantum groups $\uqgllpo$ via the Jimbo's homomorphism \cite{Jim86a}. Thus, it is very useful and interesting to study the 
Verma modules over $\uqgllpo$. In sections \ref{s:2}, \ref{s:3} and \ref{s:4} we find the explicit form of the corresponding defining relations. By this we mean the explicit expressions for the action of the generators of $\uqgllpo$ on the vectors of the natural basis of the Verma module. The corresponding representations of $\uqlsllpo$ are considered in section \ref{s:5}.

In fact, to investigate a quantum integrable system one does not need to know representations of the whole quantum loop algebra but only of its Borel subalgebras. Furthermore, the representations of the Borel subalgebras which cannot be extended to representations of the whole quantum loop algebra are of special interest. Such representations are used to construct $Q$-operators. They can be constructed as certain degeneration of the shifted Verma modules,  see, for example, \cite{BazHibKho02} and \cite{NirRaz14, BooGoeKluNirRaz14b} for the case of $\uqlslii$ and $\uqlsliii$. In the present paper we consider the general case of the quantum loop algebra $\uqlsllpo$ (section \ref{s:7}). The obtained representations appear to be reducible. We find the corresponding submodules and construct the irreducible quotient modules (section \ref{s:8}). Finally, we give an interpretation of the corresponding irreducible modules in terms of representations of the $q$-oscillator algebra (section \ref{s:9}). Almost the same expression for $q$-oscillator representations was suggested by T.~Kojima \cite{Koj08}. The advantage of our approach is that we get it as the result of degeneration of shifted Verma modules. This allows one to present $Q$-operators as a limit of transfer operators, see the papers \cite{BooGoeKluNirRaz13, BooGoeKluNirRaz14a, NirRaz14, BooGoeKluNirRaz14b} for the case $l = 1, 2$.

We assume that the deformation parameter $q \in \bbC^\times$ is not a root of unity. The notation $\kappa_q = q - q^{-1}$ is often used, so that the definition of the $q$-number can be written as 
\begin{equation*}
[\nu]_q = \frac{q^\nu - q^{- \nu}}{q - q^{-1}} 
= \kappa_q^{-1}(q^\nu - q^{-\nu}), \qquad \nu \in \bbC.
\end{equation*}
For a nonnegative integer $n$ and the corresponding $q$-number, we 
also use the notation
\begin{equation*}
[n]_q! = [1]_q \, [2]_q \, \ldots [n]_q.
\end{equation*}
It is assumed here that $[0]_q! = 1$.

\section{\texorpdfstring{Quantum group $\uqgllpo$}{Quantum group Uq(gl{l+1})}} \label{s:2}

We start with a short reminder of some basics facts on the Cartan subalgebras and root systems of the general linear and special linear Lie algebras $\gllpo$ and $\sllpo$. The standard basis of the standard Cartan subalgebra $\klpo$ of $\gllpo$ is formed by the matrices $K_i$, $i = 1, \ldots, l + 1$, with the matrix entries 
\begin{equation*}
(K_i)_{j k} = \delta_{i j} \, \delta_{i k}.
\end{equation*}
There are $l$ simple roots $\alpha_i \in \klpo^*$, which are usually defined by the equation
\begin{equation}
\langle \alpha_i, \, K_j \rangle = c_{j i}, \label{akc}
\end{equation}
where
\begin{equation}
c_{i j} = \delta_{i j} - \delta_{i, \, j + 1}. \label{cij}
\end{equation}
Then, the full system of positive roots is formed by the roots
\begin{equation*}
\alpha_{i j} = \sum_{k = i}^{j - 1} \alpha_k, \qquad 1 \le i < j \le l + 1.
\end{equation*}
It is clear that $\alpha_i = \alpha_{i, \, i + 1}$. Certainly, the negative roots are $- \alpha_{i j}$.

The standard basis of the standard Cartan subalgebra $\hlpo$ of $\sllpo$ is formed by the matrices
\begin{equation*}
H_i = K_i - K_{i + 1}, \qquad i = 1,\ldots,l.
\end{equation*}
As the positive and negative roots we take the restriction of $\alpha_{i j}$ and $- \alpha_{i j}$ to $\hlpo$. For the simple roots we have
\begin{equation}
\langle \alpha_i, \, H_j \rangle = a_{j i}, \label{aha}
\end{equation}
where
\begin{equation*}
a_{i j} = c_{i j} - c_{i + 1, \, j}
\end{equation*}
are the matrix entries of the Cartan matrix of $\sllpo$.

Let $q$ be the exponential of a complex number $\hbar$, such that $q$ is not a root of unity. We define the quantum group $\uqgllpo$ as a unital associative $\bbC$-algebra generated by the elements
\begin{equation}
E_i, \quad F_i, \quad i = 1,\ldots,l, \qquad q^X, \quad X \in \gothk_{l+1}, \label{uqglg}
\end{equation}
satisfying the following defining relations
\begin{gather}
q^0 = 1, \qquad q^{X_1} q^{X_2} = q^{X_1 + X_2}, \label{2.8a} \\
q^X E_i \, q^{-X} = q^{\langle \alpha_i, \, X \rangle} E_i, \qquad q^X F_i \, q^{-X} = q^{-\langle \alpha_i, \, X \rangle} F_i,
\label{2.9a} \\
[E_i, \, F_j] = \delta_{ij} \, \frac{q^{K_i - K_{i+1}} - q^{- K_i + K_{i + 1}}}{q - q^{-1}}.
\label{2.10a}
\end{gather}
Relations (\ref{2.9a}) can equivalently be written as
\begin{equation}
q^{\nu K_i} E_j q^{-\nu K_i} = q^{\nu c_{ij}} E_j, \qquad 
q^{\nu K_i} F_j q^{-\nu K_i} = q^{- \nu c_{ij}} F_j, \qquad \nu \in \bbC.
\label{2.11a}
\end{equation}
Besides, we have the Serre relations
\begin{gather*}
E_i E_j = E_j E_i, \qquad F_i F_j = F_j F_i, \qquad |i - j| \ge 2, \\
E_i^2 \, E_{i \pm 1} - [2]_q E_i \, E_{i \pm 1} \, E_i + E_{i \pm 1} \, E_i^2 = 0, 
\qquad 
F_i^2 \, F_{i \pm 1} - [2]_q F_i \, F_{i \pm 1} \, F_i + F_{i \pm 1} \, F_i^2 = 0.
\end{gather*}
Note that the notation $q^X$, $X \in \klpo$, is used to emphasize that the Cartan subalgebra $\klpo$ parametrizes the corresponding set of elements of $\uqgllpo$.

The quantum group $\uqsllpo$ is generated by the same generators (\ref{uqglg}) as $\uqgllpo$, however, in this case $X \in \hlpo$.  The defining relations are also the same, except that (\ref{2.10a}) should be written in the form
\begin{equation*}
[E_i, \, F_j] = \delta_{ij} \, \frac{q^{H_i} - q^{- H_i}}{q - q^{-1}}.
\end{equation*}

From the point of view of quantum integrable systems, it is important that $\uqgllpo$ and $\uqsllpo$ are Hopf algebras with respect to appropriately defined co-multiplication, antipode and counit. However, we do not use the Hopf algebra structure in the present paper.

Below we assume that
\begin{gather*}
q^{X + \nu} = q^\nu q^X, \\
[X + \nu]_q = \frac{q^{X + \nu} - q^{-X -\nu}}{q - q^{-1}} = \frac{q^{\nu} q^X - q^{-\nu} q^{-X}}{q - q^{-1}} 
\end{gather*}
for any complex number $\nu$ and element $X$ of $\klpo$.

\section{\texorpdfstring{Higher root vectors and $q$-commutation relations}{Higher root vectors and q-commutation relations}} \label{s:3}

The abelian group
\begin{equation*}
Q = \bigoplus_{i = 1}^l \bbZ \, \alpha_i
\end{equation*}
is called the root lattice of $\gllpo$. The algebra $\uqgllpo$ can be considered as $Q$-graded if we assume that
\begin{equation*}
E_i \in \uqgllpo_{\alpha_i}, \qquad F_i \in \uqgllpo_{- \alpha_i}, \qquad q^X \in \uqgllpo_0
\end{equation*}
for any $i = 1, \ldots, l$ and $X \in \klpo$. An element $a$ of $\uqgllpo$ is called a root vector corresponding to a root $\gamma$ of $\gllpo$ if $a \in \uqgllpo_\gamma$. In particular, $E_i$ and $F_i$ are root vectors corresponding to the roots $\alpha_i$ and $- \alpha_i$. It is possible to find linearly independent root vectors corresponding to all roots of $\gllpo$. To this end, we denote
\begin{equation*}
\Lambda_l = \{(i, \, j) \in \bbN \times \bbN \mid 1 \le i < j \le l + 1 \}
\end{equation*}
and, following M.~Jimbo \cite{Jim86a}, introduce elements $E_{ij}$ and $F_{ij}$, $(i, \, j) 
\in \Lambda_l$, with the help of the relations
\begin{equation*}
E_{i, \, i + 1} = E_i, \qquad i = 1, \ldots, l,
\end{equation*}
\begin{equation*}
E_{i j} = E_{i, \, j - 1} \, E_{j - 1, \, j} - q \, E_{j - 1, \, j} \, E_{i, \, j - 1}, \qquad j - i > 1,
\end{equation*} 
and
\begin{equation*} 
F_{i, \, i + 1} = F_i, \qquad i = 1,\ldots,l,
\end{equation*}
\begin{equation*}
F_{i j} = F_{j - 1, \, j} \, F_{i, \, j - 1} - q^{-1} F_{i, \, j - 1} \, F_{j - 1, \, j}, \qquad j - i > 1.
\end{equation*}
It is clear that the vectors $E_{i j}$ and $F_{i j}$ correspond to the roots $\alpha_{i j}$ and $- \alpha_{i j}$ respectively. These vectors are linearly independent, and together with the elements $q^X$, $X \in \klpo$, are called Cartan--Weyl generators of $\uqgllpo$. It appears that the ordered monomials constructed from the Cartan--Weyl generators form a Poincar\'e--Birkhoff--Witt basis of $\uqgllpo$. In this paper we choose the following ordering. First endow $\Lambda_l$ with the colexicographical order. It means that $(i, \, j) < (m, \, n)$ if $j < n$, or if $j = n$ and $i < m$.\footnote{Note that if we define an ordering of the positive roots so that $\alpha_{i j} < \alpha_{m n}$ if $(i, \, j) < (m, \, n)$ we will have a normal ordering in the sense of \cite{LezSav74, AshSmiTol79}, see also \cite{Tol89}.} Now we say that a monomial is ordered if it has the form
\begin{equation}
F_{i_1 j_1} \ldots F_{i_r j_r} \, q^X \, E_{m_1 n_1} \ldots E_{m_s n_s}, \label{fxe}
\end{equation}
where $(i_1, \, j_1) \le \ldots \le (i_r, \, j_r)$, $(m_1, \, n_1) \le \ldots \le (m_s, \, n_s)$ and $X$ is an arbitrary element of $\klpo$. The monomials of the same form with $X \in \hlpo$, form a Poincar\'e--Birkhoff--Witt basis of $\uqsllpo$.

Let us demonstrate that any monomial can be written as a sum of ordered monomials of the form (\ref{fxe}). We will write the necessary equations in the form of commutation or $q$-commutation relations. First of all, using (\ref{akc}), (\ref{aha}) and (\ref{2.11a}), we obtain
\begin{equation}
q^{\nu K_i} E_{ mn} \, q^{- \nu K_i} = q^{\nu \sum^{n - 1}_{j = m} c_{i j}} E_{m n},
\qquad
q^{\nu K_i} \, F_{m n} \, q^{- \nu K_i} = q^{- \nu \sum^{n - 1}_{j = m} c_{i j}} F_{m n},
\label{2.5}
\end{equation}
and, similarly,
\begin{equation*}
q^{\nu H_i} \, E_{m n} \, q^{- \nu H_i} = q^{\nu \sum^{n - 1}_{j = m} a_{i j}} E_{m n},
\qquad
q^{\nu H_i} \, F_{m n} \, q^{- \nu H_i} = q^{- \nu \sum^{n - 1}_{j = m} a_{i j}} F_{m n}.
\end{equation*}

To describe the relations allowing one to order the elements $E_{i j}$ and $F_{i j}$, we follow H.~Yamane \cite{Yam89}. Note that for $(i, \, j), (m, \, n) \in \Lambda_l$ such that $(i, \, j) < (m, \, n)$, there are six cases\footnote{Since we chose for the elements of $\Lambda_l$ the colexicographical order, but not the lexicographical one as in \cite{Yam89}, we define $\calC_{\mathrm{II}}$ in a different way in comparison with \cite{Yam89}.}
\begin{align}
\calC_{\mathrm I} \, : \; i = m < j < n, && \calC_{\mathrm{II}} \, : \; m < i < j < n, 
&& \calC_{\mathrm{III}} \, : \; i < m < j = n, \label{2.7} 
\\ 
\calC_{\mathrm{IV}} \, : \; i < m < j < n, && \calC_{\mathrm V} \, : \; i < j = m < n, 
&& \calC_{\mathrm{VI}} \, : \; i < j < m < n. \label{2.8}
\end{align}
Here, the symbol $\calC_a$, $a = \mathrm I, \ldots, \mathrm{VI}$, means a branch in $\Lambda_l \times \Lambda_l$,
where $(i, \, j)$ and $(m, \, n)$ are subject to the corresponding conditions (\ref{2.7})--(\ref{2.8}). In any of these branches, or in a certain union of them, the relations in question are form-invariant.

According to our definitions, which are slightly different from those of the paper \cite{Yam89}, we obtain
\begin{align}
& E_{ij} \, E_{mn} = q^{-1} E_{mn} \, E_{ij}, & & 
((i, \, j), \, (m, \, n)) \in \calC_{\mathrm I} \cup \calC_{\mathrm{III}}, \label{2.9} \\
& E_{ij} \, E_{mn} = E_{mn} \, E_{ij}, & & 
((i, \, j), \, (m, \, n)) \in \calC_{\mathrm{II}} \cup \calC_{\mathrm{VI}}, \label{2.10} \\
& E_{ij} \, E_{mn} - q \, E_{mn} \, E_{ij} = E_{in}, & & 
((i, \, j), \, (m, \, n)) \in \calC_{\mathrm{V}}, \label{2.11} \\
& E_{ij} \, E_{mn} - E_{mn} \, E_{ij} = -\kappa_q \, E_{in} \, E_{mj}, & & 
((i, \, j), \, (m, \, n)) \in \calC_{\mathrm{IV}}. \label{2.12} 
\end{align}
Similarly, we have
\begin{align}
& F_{ij} \, F_{mn} = q^{-1} F_{mn} \, F_{ij}, & & 
((i, \, j), \, (m, \, n)) \in \calC_{\mathrm I} \cup \calC_{\mathrm{III}}, \label{2.13} \\
& F_{ij} \, F_{mn} = F_{mn} \, F_{ij}, & & 
((i, \, j), \, (m, \, n)) \in \calC_{\mathrm{II}} \cup \calC_{\mathrm{VI}}, \label{2.14} \\
& F_{ij} \, F_{mn} - q \, F_{mn} \, F_{ij} = -q \, F_{in}, & & 
((i, \, j), \, (m, \, n)) \in \calC_{\mathrm{V}}, \label{2.15} \\
& F_{ij} \, F_{mn} - F_{mn} \, F_{ij} = -\kappa_q \, F_{in} \, F_{mj}, & & 
((i, \, j), \, (m, \, n)) \in \calC_{\mathrm{IV}}. \label{2.16} 
\end{align}
It is convenient to write (\ref{2.15}) also as
\begin{equation*}
F_{mn} \, F_{ij} - q^{-1} F_{ij} \, F_{mn} = F_{in}.
\end{equation*}

Further, we obtain
\begin{equation*}
[E_{ij} \, , \, F_{ij}] = \kappa_q^{-1} 
\big( q^{H_{i j}} - 
 q^{- H_{i j}} \big),
\end{equation*}
where, and in what follows, $[ \; , \, ]$ means the usual commutator, and we denote
\begin{equation*}
q^{\nu H_{i j}} = q^{\nu \sum_{k = i}^{j - 1} H_k}, \qquad \nu \in \bbC, \qquad (i, \, j) \in \Lambda_l.
\end{equation*}
We also obtain the following commutation relations:
\begin{align}
& [E_{ij} \, , \, F_{mn}] = - q^{-1} F_{jn} \, q^{- H_{i j}}, & & 
((i, \, j), \, (m, \, n)) \in \calC_{\mathrm I}, \label{2.18} \\
& [E_{ij} \, , F_{mn}] = q \, E_{im} \, q^{- H_{m n}} = 
q^{- H_{m n}} E_{im},& &  
((i, \, j), \, (m, \, n)) \in \calC_{\mathrm{III}}, \label{2.19} \\
& [E_{ij} \, , \, F_{mn}] = 0, & &
((i, \, j), \, (m, \, n)) \in \calC_{\mathrm{II}} \cup \calC_V \cup \calC_{\mathrm{VI}}, 
\label{2.20} \\
& [E_{ij} \, , \, F_{mn}] = \kappa_q \, F_{j n} \, E_{im} \, q^{- H_{m j}},
& &  ((i, \, j), \, (m, \, n)) \in \calC_{\mathrm{IV}}. \label{2.21}
\end{align}
Note that in (\ref{2.21}) the root vectors at the right hand side commute, $[F_{j n}, \, E_{i m}] = 0$ for the given values of the indices, and, besides,
\begin{equation*}
q^{- H_{m j}} \, F_{jn} = q^{-1} F_{jn} \, q^{- H_{m j}},
\qquad \qquad
q^{- H_{m j}} \, E_{im} = q \, E_{im} \, q^{- H_{m j}}.
\end{equation*}
Interchanging the pairs of indices $(i, \, j)$ and $(m, \, n)$ of the root vectors in the above commutation relations, additionally to (\ref{2.18})--(\ref{2.21}) we obtain 
\begin{align}
& [E_{m n} \, , \, F_{i j}] = - q \, q^{H_{i j}} \, E_{jn} = 
- E_{jn} \, q^{H_{i j}}, & & 
((i, \, j), \, (m, \, n)) \in \calC_{\mathrm I}, \label{2.22} \\
& [E_{mn} \, , F_{ij}] = q^{-1} q^{H_{m n}} \, F_{im} = 
F_{im} \, q^{H_{m n}}, & & 
((i, \, j), \, (m, \, n)) \in \calC_{\mathrm{III}}, \label{2.23} \\
& [E_{mn} \, , \, F_{ij}] = 0, & &
((i, \, j), \, (m, \, n)) \in \calC_{\mathrm{II}} \cup \calC_V \cup \calC_{\mathrm{VI}}, 
\label{2.24}  \\
& [E_{mn} \, , \, F_{ij}] = - \kappa_q \, F_{im} \, E_{jn} \, q^{H_{m j}},
& & ((i, \, j), \, (m, \, n)) \in \calC_{\mathrm{IV}}. \label{2.25} 
\end{align}
Again, the root vectors at the right hand side of (\ref{2.25}) commute for the given values of the indices, $[F_{i m}, \, E_{j n}] = 0$, and similarly to the preceding we have
\begin{equation*}
q^{H_{m j}} \, E_{jn} = q^{-1} E_{jn} \, q^{H_{m j}},
\qquad \qquad
q^{H_{m j}} \, F_{im} = q \, F_{im} \, q^{H_{m j}}.
\end{equation*}

Now, it is easy to demonstrate that equations (\ref{2.5}) and (\ref{2.9})--(\ref{2.25}) are sufficient to rewrite any monomial as a sum of ordered monomials of the form (\ref{fxe}). In the case of the quantum group $\uqsllpo$ we obtain the same result using the ordered monomials of the form (\ref{fxe}) with $X \in \hlpo$.

\section{\texorpdfstring{Defining Verma $\uqgllpo$-module relations}{Defining Verma Uq(gl(l+1))-module relations}} \label{s:4}

Given $\lambda \in \klpo^*$, denote by $\widetilde V^\lambda$ the corresponding Verma $\uqgllpo$-module. This is a highest weight module with the highest weight vector $v^\lambda$ satisfying the relations
\begin{equation}
E_i \, v^\lambda = 0, \quad i = 1,\ldots,l, \qquad 
q^X v^\lambda = q^{\langle \lambda, \, X \rangle} v^\lambda,
\quad X \in \gothk_{l+1}, \quad \lambda \in \gothk_{l+1}^*. \label{3.1}
\end{equation}
Below we identify the highest weight $\lambda$ with the set of its components
\begin{equation*}
\lambda_i = \langle \lambda, \, K_i \rangle.
\end{equation*}
We denote by $\widetilde \pi^\lambda$ the representation of $\uqgllpo$ corresponding 
to $\widetilde V^\lambda$. The structure and properties of $\widetilde V^\lambda$ and 
$\widetilde \pi^\lambda$ for $l = 1$ and $l = 2$ are considered in much detail in our papers \cite{BooGoeKluNirRaz13, BooGoeKluNirRaz14a, NirRaz14, BooGoeKluNirRaz14b, BooGoeKluNirRaz15}. Here we deal with the case of general $l$.

Denote by $\bm m$ the $l (l + 1)/2$-tuple of non-negative integers $m_{i j}$, $(i, \, j) \in \Lambda_l$, arranged in the colexicographical order of $(i, \, j)$. More explicitly,
\begin{equation*}
{\bm m} = (m_{12}, \, m_{13}, \, m_{23}, \, \ldots, \, m_{1j}, \, \ldots, \, m_{j-1, \, j}, \, \ldots, \,
m_{1, \, l + 1}, \, \ldots, \, m_{l, \, l + 1}).
\end{equation*}
The vectors
\begin{equation*}
v_{\bm m} = F_{12}^{m_{12}} \, F_{13}^{m_{13}} \, F_{23}^{m_{23}} \, 
\ldots F_{1, \, j}^{m_{1, \, j}} \ldots F_{j - 1, \, j}^{m_{j - 1, \, j}} \ldots 
F_{1, \, l + 1}^{m_{1, \, l + 1}} \ldots F_{l, \, l + 1}^{m_{l, \, l + 1}} \, v_{\bf0},
\end{equation*}
where for consistency we denote $v_{\bm 0} = v^\lambda$, form a basis of $\widetilde V^\lambda$. The relations describing the action of the generators $q^{\nu K_i}$, $E_i$ and $F_i$ of the quantum group $\uqgllpo$ on a general basis vector $v_{\bm m}$ is what we exactly mean under defining $\uqgllpo$-module relations.

We first obtain how the generators $q^{\nu K_i}$ act on the basis vectors. Using (\ref{2.5}) 
and taking into account the second relation of (\ref{3.1}), we derive
\begin{equation*}
q^{\nu K_i} v_{\bm m} = q^{\nu (\lambda_i - \sum_{j=1}^l \sum_{r=1}^j
\sum_{s=j+1}^{l+1} c_{ij} m_{rs})} v_{\bm m}, \qquad i = 1,\ldots,l+1.
\end{equation*}
Rearranging the summations in the exponential at the right hand side of
the above equation, the same result can be written as
\begin{equation*}
q^{\nu K_i} v_{\bm m} = q^{\nu (\lambda_i - \sum_{s=2}^{l+1} \sum_{r=1}^{s-1}
\sum_{j=r}^{s-1} m_{rs} c_{ij})} v_{\bm m}, \qquad i = 1,\ldots,l+1.
\end{equation*}
Recalling the exact form (\ref{cij}) of the quantities $c_{ij}$, we can make the above formulas more explicit and eligible for further use. We
have
\begin{equation}
q^{\nu K_i} \, v_{\bm m} = q^{\nu (\lambda_i + \sum_{k=1}^{i-1} m_{ki}
- \sum_{k=i+1}^{l+1} m_{ik})} v_{\bm m}, \qquad i = 1,\ldots,l+1.
\label{3.3}
\end{equation} 
Then, for $q^{\nu H_i}$, $i = 1,\ldots,l$, where $H_i = K_i - K_{i+1}$, we obtain
\begin{equation}
q^{\nu H_i} \, v_{\bm m} = q^{\nu [\lambda_i - \lambda_{i+1} 
+ \sum_{k=1}^{i-1} (m_{ki} - m_{k,i+1}) - 2 m_{i,i+1}  
- \sum_{k=i+2}^{l+1} (m_{ik} - m_{i+1,k})]} v_{\bm m}.
\label{3.3x}
\end{equation} 

To define the action of the generators $F_k = F_{k, \, k+1}$ on the basis vectors we need subsidiary formulas following from relations (\ref{2.13})--(\ref{2.16}). These are
\begin{equation*}
F_{k, \, k + 1} \, F_{i k}^{m_{i k}} = q^{- m_{i k}} \, F_{i k}^{m_{i k}} \, F_{k, \, k+1} + [m_{i k}]_q \, F_{i k}^{m_{i k} - 1} \, F_{i, \,k+1}
\end{equation*} 
for $i = 1, \ldots, k - 1$ and $k = 2, \ldots, l$, and
\begin{equation*}
F_{j, \, k + 1}^\ms \, F_{i, \, k + 1}^{m_{i, \, k + 1}} = q^{m_{i, \, k + 1}} \, F_{i, \, k + 1}^{m_{i, \, k + 1}} \, F_{j, \, k + 1}^\ms, \qquad i < j = 2, \ldots, k.
\end{equation*} 
Applying these formulas for all $k = 1,\ldots,l$, we obtain
\begin{equation}
F_{i, \, i + 1} \, v_{\bm m} = q^{- \sum_{k = 1}^{i-1} (m_{k i} - m_{k, \, i + 1})} \, v_{{\bm m} + \epsilon_{i, \, i + 1}} + \sum_{j = 1}^{i - 1} q^{- \sum_{k=1}^{j - 1} (m_{k i} - m_{k, \, i+1})} \, [m_{j i}]_q \, v_{{\bm m} - \epsilon_{j i} + \epsilon_{j, \, i + 1}}.
\label{3.6}
\end{equation}
Here and below ${\bm m} + \nu \epsilon_{i j}$ means shifting by $\nu$ the entry $m_{i j}$ in the $l(l + 1)/2$-tuple ${\bm m}$.

To define the action of the generators $E_k = E_{k, \, k + 1}$, $k = 1, \ldots, l$, on the basis vectors we mainly need the following subsidiary formulas obtained from equations (\ref{2.18})--(\ref{2.21}) and (\ref{2.22})--(\ref{2.25}):
\begin{align*}
E_{k, \, k + 1}^\ms \, F_{k, \, k + 1}^{m_{k, \, k + 1}} & = F_{k, \, k + 1}^{m_{k, \, k + 1}} \, E_{k, \, k + 1}^\ms + [m_{k, \, k + 1}]_q \, F_{k, \, k + 1}^{m_{k, \, k + 1}-1} \, [H_k - m_{k, \, k + 1} + 1]_q, \\
E_{k, \, k + 1}^\ms \, F_{k j}^{m_{k j}} & = F_{k j}^{m_{k j}} \, E_{k, \, k+1}^\ms - q^{m_{k j} - 2} \, [m_{k j}]_q \, F_{k j}^{m_{k j} - 1} \, F_{k + 1, \, j}^\ms \, q^{-H_k},
\\
E_{k, \, k + 1} \, F_{i, \, k + 1}^{m_{i, \, k + 1}} & = F_{i,\, k + 1}^{m_{i, \, k + 1}} \, E_{k, \, k + 1} +
[m_{i, \, k + 1}]_q \, F_{i k} \, F_{i, \, k + 1}^{m_{i, \, k + 1} - 1} \, q^{H_k}
\end{align*}
for all possible values of $i$, $j$ and $k$. 
These equations, supplied with (\ref{3.3x}), allow us to derive the desirable formula. We obtain
\begin{align}
& E_{i, \, i + 1} \, v_{\bm m} = [\lambda_i - \lambda_{i+1} - \sum_{j = i + 2}^{l + 1} (m_{i j} - m_{i + 1, \, j}) 
- m_{i, \, i + 1} + 1]_q \, [m_{i, \, i + 1}]_q \, v_{{\bm m} - \epsilon_{i, \, i + 1}} \notag \\*
& \hspace{1.5em} {} + q^{\lambda_i - \lambda_{i + 1} - 2 m_{i, \, i + 1} 
- \sum_{j = i + 2}^{l + 1} (m_{i j} - m_{i + 1, \, j})} 
\sum_{j = 1}^{i - 1} q^{\sum_{k = j + 1}^{i - 1} (m_{k i} - m_{k,\, i + 1})} \,
[m_{j,\, i + 1}]_q \, v_{{\bm m} - \epsilon_{j, \, i + 1}  + \epsilon_{j i}} \notag \\
& \hspace{11.5em} {} - \sum_{j = i + 2}^{l + 1} q^{- \lambda_i + \lambda_{i + 1} - 2 + \sum_{k = j}^{l + 1} 
(m_{i k} - m_{i + 1, \, k})} \, [m_{i j}]_q \, v_{{\bm m} - \epsilon_{i j} + \epsilon_{i + 1, \, j}} 
\label{3.8}
\end{align}
for all $i = 1, \ldots, l$. 

To construct representations of $\uqlsllpo$ we will also need in section \ref{s:5} the action of the specific root vectors $F_{1, \, l+1}$ and $E_{1, \, l+1}$ on the basis vectors $v_{\bm m}$. Using the formulas
\begin{equation*}
F_{1, \, l + 1} \, F_{1 j}^{m_{1 j}} = q^{m_{1 j}} \, F_{1 j}^{m_{1 j}} \, F_{1, \, l + 1}, \qquad j = 2,\ldots,l,
\end{equation*}
and
\begin{equation*}
F_{1, \, l + 1} \, F_{i j}^{m_{i j}} = F_{i j}^{m_{i j}} \, F_{1, \, l+1}, \qquad (i, \, j) \in \Lambda_l,
\end{equation*}
following from equations (\ref{2.13})--({\ref{2.16}}), we obtain
\begin{equation}
F_{1, l + 1} \, v_{\bm m} = q^{\sum_{i = 2}^l m_{1 i}} \, v_{{\bm m} + \epsilon_{1, \, l + 1}}.
\label{3.7}
\end{equation}
The corresponding formula for the action of $E_{1, \, l + 1}$ is given in the appendix.

Note that $\widetilde V^\lambda$ and $\widetilde \pi^\lambda$ are infinite dimensional for the general weights $\lambda \in \gothk^*_{l+1}$. However, if the weights are such that $\lambda_i - \lambda_{i+1} \in \bbZ_+$ for all $i = 1,\ldots,l$, there is a maximal submodule, such that the respective quotient module is finite dimensional. We denote such $\uqgllpo$-module and the corresponding representation by $V^\lambda$ and $\pi^\lambda$, respectively. The reduction to the special linear case from the general linear one can be achieved simply by replacing relation (\ref{3.3}) by (\ref{3.3x}) in the above module relations.   

\section{\texorpdfstring{Quantum loop algebras $\uqlsllpo$ and some their representations}{Quantum loop algebras Uq(L(sl(l+1))) and some their representations}} 
\label{s:5}

For the construction and investigation of quantum integrable systems one often uses finite and infinite dimensional representations of the quantum loop algebra $\uqlsllpo$. The relevant representations are usually obtained from Verma representations of the quantum group $\uqgllpo$. Let us describe the corresponding procedure. 

To define the quantum loop algebra $\uqlsllpo$ it is convenient to start with the definition of the quantum group $\uqtlsllpo$. Remind that the Lie algebra $\tlsllpo$ is a special extension of the loop algebra $\lsllpo$ by a one-dimensional center $\bbC c$ \cite{Kac90}. For any $i = 1, \ldots, l$ denote by $h_i$ the image of the Cartan elements $H_i$ of $\sllpo$ under the natural embedding of $\sllpo$ into $\tlsllpo$. We will also use the notation
\begin{equation*}
\thlpo = \bbC c \oplus \bigoplus_{i = 1}^l \bbC h_i
\end{equation*}
for the `Cartan subalgebra' of $\tlsllpo$. Introducing the element $h_0 = c - \sum_{i = 1}^l h_i$, we obtain a more symmetric expression:
\begin{equation*}
\thlpo = \bigoplus_{i = 0}^l \bbC h_i.
\end{equation*}
We define the `simple roots' $\alpha_i$, $i = 0, 1, \ldots, l$, of $\tlsllpo$ as the elements of $\thlpo^*$ satisfying the equation
\begin{equation*}
\langle \alpha_i, \, h_j \rangle = \widetilde a_{j i},
\end{equation*}
where $\widetilde a_{ij}$ are the entries of the generalized Cartan matrix of an affine Lie algebra of type~$A^{(1)}_{l}$.

The quantum group $\uqtlsllpo$ is a unital associative $\bbC$-algebra generated by the elements $e_i$, $f_i$, $i = 0, 1, \ldots, l$, and $q^x$, $x \in \widetilde \gothh_{l+1}$, satisfying certain defining relations. These are the following commutation relations:
\begin{gather*}
q^0 = 1, \qquad q^{x_1} q^{x_2} = q^{x_1 + x_2}, \\
q^x e_i \, q^{-x} = q^{\langle \alpha_i, \, x \rangle} e_i, \qquad q^x f_i \, q^{-x} = q^{- \langle \alpha_i, \, x \rangle} f_i, \\
[e_i, \, f_j] = \delta_{ij} \, \frac{q^{h_i} - q^{-h_i}}{q - q^{-1}},
\end{gather*}
supplemented by the Serre relations:
\begin{equation*}
\sum_{k = 0}^{1 - \widetilde a_{ij}} (-1)^k  
(e_i)^{(1 - \widetilde a_{ij} - k)} \, e_j \, (e_i)^{(k)} = 0, 
\qquad
\sum_{k = 0}^{1 - \widetilde a_{ij}} (-1)^k  
(f_i)^{(1 - \widetilde a_{ij} - k)} \, f_j \, (f_i)^{(k)} = 0.
\end{equation*}
Here we use the notation $(e_i)^{(n)} = (e_i)^n / [n]_q!$ and $(f_i)^{(n)} = (f_i)^n / [n]_q!$.

The quantum group $\uqtlsllpo$ does not have any finite dimensional representations with $q^{\nu c}$ acting nontrivially \cite{ChaPre91, ChaPre94}. Therefore, to construct quantum integrable systems with finite dimensional state space one should use representations with trivial action of $q^{\nu c}$. In fact, it appears more convenient  to use the quantum loop algebra $\uqlsllpo$ defined as the quotient
\begin{equation*}
\uqlsllpo = \uqtlsllpo / \langle q^{\nu c} - 1 \rangle_{\nu \in \bbC}.
\end{equation*}
We consider the quantum loop algebra $\uqlsllpo$ in terms of the same generators and defining relations as $\uqtlsllpo$, with the additional relations
\begin{equation}
q^{\nu c} = 1, \qquad \nu \in \bbC^\times. \label{qnuc}
\end{equation}
As the quantum group $\uqgllpo$, also the quantum loop algebra $\uqlsllpo$ is a Hopf algebra with respect to appropriately defined co-multiplication, antipode and counit.

To construct representations of $\uqlsllpo$, not only finite dimensional but also infinite
dimensional ones, it is common to use the Jimbo's homomorphism $\varepsilon$ from the quantum loop algebra $\uqlsllpo$ to the quantum group $\uqgllpo$
defined by the equations \cite{Jim86a}
\begin{align}
& \varepsilon(q^{\nu h_0}) = q^{\nu (K_{l+1} - K_1)}, &&  
 \varepsilon(q^{\nu h_i}) = q^{\nu (K_{i} - K_{i+1})}, \label{4.9}\\
& \varepsilon(e_0) = F_{1,\, l+1} \, q^{K_1 + K_{l+1}}, && 
 \varepsilon(e_i) = E_{i, \,i+1}, \label{4.10} \\
& \varepsilon(f_0) = E_{1, \, l+1} \, q^{-K_1 - K_{l+1}}, && 
  \varepsilon(f_i) = F_{i, \, i+1}, \label{4.11}  
\end{align}
where $i$ runs from $1$ to $l$. If $\pi$ is a representation of $\uqgllpo$ then $\pi \circ \varepsilon$ is a representation of $\uqlsllpo$.

In fact, in applications to the theory of quantum integrable systems, one usually considers families of representations parametrized by a complex parameter called a spectral parameter. We introduce a spectral parameter with the help of a family of mappings $\Gamma_\zeta \colon\uqlsllpo \to \uqlsllpo$, $\zeta \in \bbC^\times$. Explicitly, $\Gamma_\zeta$ is defined by its action on the generators as
\begin{equation*}
\Gamma_\zeta(q^x) = q^x, \qquad \Gamma_\zeta(e_i) = \zeta^{s_i} e_i, \qquad \Gamma_\zeta(f_i) = \zeta^{-s_i} f_i,
\end{equation*}
where $s_i$ are arbitrary integers. We denote the total sum of these integers by $s$. Now, for any representation $\varphi$ of $\uqlsllpo$ we define the corresponding family $\varphi_\zeta$ of representations as
\begin{equation*}
\varphi_\zeta = \varphi \circ \Gamma_\zeta.
\end{equation*}
Of our special interest are the families of representations $(\widetilde \varphi^\lambda)_\zeta$ and $(\varphi^\lambda)_\zeta$ related to infinite and finite dimensional representations $\widetilde \pi^\lambda$ and $\pi^\lambda$ of $\uqgllpo$, because they play a special role in the theory of quantum integrable systems. These families are defined as
\begin{equation*}
(\widetilde \varphi{}^\lambda)_\zeta  = \widetilde \pi^\lambda \circ \varepsilon \circ \Gamma_\zeta, \qquad (\varphi^\lambda)_\zeta = \pi^\lambda \circ \varepsilon \circ \Gamma_\zeta.
\end{equation*}
Let us consider the corresponding defining $\uqlsllpo$-module relations.

First, using (\ref{3.3}) and (\ref{4.9}), we obtain
\begin{align}
q^{\nu h_0} \, v_{\bm m} & = q^{\nu [\lambda_{l + 1} - \lambda_1 + \sum_{i = 2}^l (m_{1 i} + m_{i, \, l + 1}) + 2 m_{1, \, l + 1}]} \, v_{\bm m}, 
\label{4.12} \\
q^{\nu h_i} \, v_{\bm m} & = q^{\nu [\lambda_i - \lambda_{i + 1} + \sum_{k = 1}^{i - 1} (m_{k i} - m_{k, \, i + 1}) - 2 m_{i, \, i+1} - \sum_{k = i + 2}^{l + 1} (m_{i k} - m_{i + 1, \, k})]} v_{\bm m}.
\label{4.13}
\end{align}

Further, taking into account that
\begin{equation*}
q^{K_1 + K_{l + 1}} \, v_{\bm m} = q^{\lambda_1 + \lambda_{l + 1} - \sum_{i = 2}^l (m_{1 i} - m_{i, \, l + 1})} \, v_{\bm m}
\end{equation*}
and using (\ref{3.7}), we obtain from the first equation of (\ref{4.10}) the module relation for $e_0$,
\begin{equation}
e_0 \, v_{\bm m} = \zeta^{s_0} \, q^{\lambda_1 + \lambda_{l + 1} + \sum_{i = 2}^l m_{i, \, l+1}} \, v_{{\bm m} + \epsilon_{1, \, l + 1}}.
\label{4.14}
\end{equation}
The module relations for $e_i$, $i = 1,\ldots,l$, follow from the second equation of (\ref{4.10}) with account of equation (\ref{3.8}),
\begin{align}
& e_i \, v_{\bm m} = \zeta^{s_i} [\lambda_i - \lambda_{i+1} - \sum_{j = i + 2}^{l + 1} (m_{i j} - m_{i + 1, \, j}) 
- m_{i, \, i + 1} + 1]_q \, [m_{i, \, i + 1}]_q \, v_{{\bm m} - \epsilon_{i, \, i + 1}} \notag \\*
& \hspace{1.5em} {} + \zeta^{s_i} q^{\lambda_i - \lambda_{i + 1} - 2 m_{i, \, i + 1} 
- \sum_{j = i + 2}^{l + 1} (m_{i j} - m_{i + 1, \, j})} 
\sum_{j = 1}^{i - 1} q^{\sum_{k = j + 1}^{i - 1} (m_{k i} - m_{k,\, i + 1})} \,
[m_{j,\, i + 1}]_q \, v_{{\bm m} - \epsilon_{j, \, i + 1}  + \epsilon_{j i}} \notag \\
& \hspace{11.5em} {} - \zeta^{s_i} \sum_{j = i + 2}^{l + 1} q^{- \lambda_i + \lambda_{i + 1} - 2 + \sum_{k = j}^{l + 1} (m_{i k} - m_{i + 1, \, k})} \, [m_{i j}]_q \, v_{{\bm m} - \epsilon_{i j} + \epsilon_{i + 1, \, j}}. 
\label{4.15}
\end{align}

Using (\ref{3.6}) and the second relation from (\ref{4.11}), we obtain the action of the generators $f_i$ on the basis vectors $v_{\bmm}$:
\begin{multline}
f_i \, v_{\bmm} = \zeta^{-s_i} \, q^{- \sum_{j = 1}^{i - 1} (m_{j i} - m_{j, \, i + 1})} \, 
v_{{\bmm} + \epsilon_{i, \, i + 1}} \\* + \zeta^{-s_i} \, \sum_{j=1}^{i-1} 
q^{- \sum_{k = 1}^{j-1} (m_{k i} - m_{k, \, i + 1})} \, [m_{ji}]_q \, 
v_{{\bmm} - \epsilon_{j i} + \epsilon_{j, \, i + 1}}. 
\label{4.15f}
\end{multline}
One can also obtain the action of $f_0$ on $v_{\bmm}$ using equations (\ref{E1l1}), (\ref{3.3}) and the first one of the maps given by (\ref{4.11}). 

Now, relations (\ref{4.12})--(\ref{4.13}), (\ref{4.14})--(\ref{4.15}), and (\ref{4.15f}) together with an expression for $f_0 v_{\bmm}$ constructed as just mentioned above, form the basic $\uqlsllpo$-module relations.

\section{\texorpdfstring{Degenerations of the shifted $\uqbp$-modules}{Degenerations of the shifted Uq(b+)-modules}} \label{s:7}

From the point of view of quantum integrable systems, representations of the Borel subalgebras of the quantum groups under consideration are most interesting. There are two standard Borel subalgebras of $\uqlsllpo$, the positive Borel subalgebra $\uqbp$ generated by $e_i$, $i = 0, 1, \ldots, l$ and $q^x$, $x \in \thlpo$, and the negative Borel subalgebra $\uqbm$ generated by $f_i$, $i = 0, 1, \ldots, l$ and $q^x$, $x \in \thlpo$. We restrict ourselves to the case of the positive Borel subalgebra used in our consideration of universal integrability objects \cite{BooGoeKluNirRaz13, BooGoeKluNirRaz14a, BooGoeKluNirRaz14b, NirRaz14}.

Certainly, the restriction of any representation of $\uqlsllpo$ to $\uqbp$ is a representation of $\uqbp$. In particular, we can consider the restriction of the representations $(\widetilde \varphi^\lambda)_\zeta$ and $(\varphi^\lambda)_\zeta$. Here relations (\ref{4.12})--(\ref{4.13}) and (\ref{4.14})--(\ref{4.15}) 
constitute the corresponding $\uqbp$-module relations. The representations  $(\widetilde \varphi^\lambda)_\zeta$ and $(\varphi^\lambda)_\zeta$ are used for the construction of very important integrability objects called transfer operators. Besides, there are no less important integrability objects called $Q$-operators. The representations used for the construction of the $Q$-operators are essentially different. However, for the quantum integrable systems related to the quantum groups $\uqlsllpo$ the latter can be obtained from the former as certain degenerations, see, for example, \cite{BazHibKho02} and \cite{NirRaz14, BooGoeKluNirRaz14b}.

The degenerations in question are obtained by sending each difference $\lambda_i - \lambda_{i + 1}$, $i = 1, \ldots, l$, to positive or negative infinity. However, looking at (\ref{4.12}) and (\ref{4.13}) we see that it gives either infinity or zero for the action of the corresponding elements $q^{\nu h_i}$. To overcome this difficulty we use the notion of a shifted representation.  

Let $\varphi$ be a representation of $\uqbp$ and $\xi \in \thlpo^*$. Then the relations
\begin{equation}
\varphi[\xi](e_i) = \varphi(e_i), \qquad \varphi[\xi](q^x) = q^{\langle \xi, \, x \rangle} \varphi(q^x) 
\label{shift}
\end{equation}
define a representation $\varphi[\xi]$ of $\uqbp$ called a shifted representation. Note that due to (\ref{qnuc}) the element $\xi$ must satisfy the equation
\begin{equation*}
\langle \xi, \, c \rangle = 0.
\end{equation*}
We see that the only difference between the shifted and initial representations appears in the factor $q^{\langle \xi, \, x \rangle}$. 

Now we consider the $\uqbp$-module defined by relations (\ref{4.12})--(\ref{4.13}) and (\ref{4.14})--(\ref{4.15}), and perform there a shift according to definition (\ref{shift}), with $\xi$ specified by the equations
\begin{equation*}
\langle \xi, \, h_0 \rangle = - \lambda_{l + 1} + \lambda_1, \qquad 
\langle \xi, \, h_i \rangle = -\lambda_i + \lambda_{i + 1},
\qquad i = 1,\ldots, l.
\end{equation*}
Obviously, this shift has an effect only on the relations describing the action of the generators $q^{\nu h_i}$, $i = 0, 1, \ldots, l$. Now we have
\begin{align*}
q^{\nu h_0} \, v_{\bm m} & = q^{\nu (\sum_{j = 2}^l (m_{1 j} + m_{j, \, l + 1}) + 2 m_{1,\, l + 1})} \, v_{\bm m}, \\
q^{\nu h_i} \, v_{\bm m} & = q^{\nu (\sum_{j = 1}^{i - 1} (m_{j i} - m_{j, \, i + 1}) - 2 m_{i, \, i + 1} - \sum_{j = i + 2}^{l + 1} (m_{i j} - m_{i + 1, \, j}))} v_{\bm m}.
\end{align*}
Since these relations do not contain $\lambda_i$, $i = 1, \ldots, l + 1$, we can allow the differences $\lambda_i - \lambda_{i + 1}$, $i = 1, \ldots, l$, to go in these relations to positive or negative infinity. To be concrete, we are going to consider the limit
\begin{equation}
\lambda_i - \lambda_{i + 1} \to - \infty, \qquad i = 1, \ldots, l. \label{lilipo}
\end{equation}
The representations which can be obtained with other choices can be obtained then with the help of automorphisms of $\uqbp$.

It is clear that some problems with the limit (\ref{lilipo}) for relations (\ref{4.14}) and (\ref{4.15}) remain. Let us define a new basis in the representation space formed by the vectors
\begin{equation*}
w_{\bm m} = c_{\bm m} v_{\bm m},
\end{equation*}
where
\begin{equation*}
c_{\bm m} = q^{\sum_{i = 1}^ l [\lambda_i - \lambda_{i + 1} + 1 + (2 \lambda_{l + 1} - l) s_i / s] \sum_{j = 1}^i \sum_{k = i + 1}^{l + 1} m_{j k}}.
\end{equation*}
Here we note that
\begin{equation*}
c_{\bm m + \nu \epsilon_{i j}} 
=  q^{\nu \sum_{k = i}^{j - 1} [\lambda_k - \lambda_{k + 1} + 1 + (2 \lambda_{l + 1} - l) s_k / s]}  c_{\bm m}.
\end{equation*}
Recall that $s$ denotes the total sum of the integers $s_i$, $i = 0, 1, \ldots, l$. Using this equation, we obtain from (\ref{4.14})--(\ref{4.15}) the following $\uqbp$-module relations in the new basis,
\begin{equation*}
e_0 \, w_{\bm m} = \widetilde\zeta^{s_0} \, q^{\sum_{i = 2}^l m_{i, \, l + 1}} \, 
w_{{\bm m} + \epsilon_{1, \, l + 1}}
\end{equation*}
and
\begin{align*}
e_i \, w_{\bm m} & = \widetilde\zeta^{s_i} \, \kappa_q^{-1} \, 
\Big( q^{2(\lambda_i - \lambda_{i + 1}) + 2 - 
\sum_{j = i + 2}^{l + 1} (m_{i j} - m_{i + 1, \, j}) - m_{i, \, i + 1}} \notag \\
& \hspace{6.5em} {} - q^{\sum_{j = i + 2}^{l + 1} (m_{i j} - m_{i + 1, \, j}) + m_{i, \, i + 1}} \Big) [m_{i, \, i + 1}]_q \, w_{{\bm m} - \epsilon_{i, \, i + 1}} \notag \\
& + \widetilde\zeta^{s_i} \, q^{2(\lambda_i - \lambda_{i + 1}) + 1 - 2 m_{i, \, i + 1} - \sum_{j = i + 2}^{l + 1} (m_{i j} - m_{i + 1, \, j})} \notag \\
& \hspace{8.1em} \times \sum_{j = 1}^{i - 1} q^{\sum_{k = j + 1}^{i - 1} (m_{k i} - m_{k, \, i + 1})} \,
[m_{j, \, i + 1}]_q \, w_{{\bm m} - \epsilon_{j, \, i + 1} + \epsilon_{j i}} \notag \\
& \hspace{7em} {} - \widetilde\zeta^{s_i} \, \sum_{j = i + 2}^{l + 1} q^{\sum_{k = j}^{l + 1} (m_{i k} - m_{i + 1, \, k}) - 1} \, [m_{i j}]_q \, w_{{\bm m} - \epsilon_{i j} + \epsilon_{i + 1, \, j}},
\end{align*}
where $\widetilde\zeta$ is the new spectral parameter defined as
\begin{equation*}
\widetilde\zeta = q^{(2\lambda_{l+1} - l)/s} \, \zeta.
\end{equation*}

Now we can consider the infinite limit (\ref{lilipo}). The final result is a degeneration of the shifted $\uqbp$-module described by the relations
\begin{align}
q^{\nu h_0} \, w_{\bm m} & = q^{\nu (\sum_{i = 2}^l (m_{1 i} + m_{i, \, l+1}) + 2 m_{1,\, l + 1})} \, w_{\bm m}, 
\label{5.4} \\
q^{\nu h_i} \, w_{\bm m} & = q^{\nu (\sum_{j = 1}^{i - 1} (m_{j i} - m_{j, \, i + 1}) - 2 m_{i, \, i + 1}  
- \sum_{j = i + 2}^{l + 1} (m_{i j} - m_{i + 1, \, j}))} w_{\bm m}
\label{5.5}
\end{align}
and
\begin{align}
e_0 \, w_{\bm m} & = \widetilde\zeta^{s_0} \, q^{\sum_{i = 2}^l m_{i, \, l + 1}} \, 
w_{{\bm m} + \epsilon_{1, \, l + 1}}, \label{5.6} \\*
e_i \, w_{\bm m} & = {} - \widetilde\zeta^{s_i} \, \kappa_q^{-1} \, q^{\sum_{j = i + 2}^{l + 1} (m_{i j} - m_{i + 1, \, j}) + m_{i, \, i + 1}} [m_{i, \, i + 1}]_q \, w_{{\bm m} - \epsilon_{i, \, i + 1}} \notag \\*
& \hspace{8em} {} - \widetilde\zeta^{s_i} \, \sum_{j = i + 2}^{l + 1} q^{\sum_{k = j}^{l + 1} (m_{i k} - m_{i + 1, \, k}) - 1} \, [m_{i j}]_q \, w_{{\bm m} - \epsilon_{i j} + \epsilon_{i + 1, \, j}}. \label{5.7}
\end{align}
These are our main $\uqbp$-module relations on the basis of which we will make all the subsequent constructions.

\section{Factoring out the submodules}
\label{s:8}

We denote by $\rho''$ the representation of $\uqbp$ defined by the relations
\begin{align*}
q^{\nu h_0} \, v_{\bm m} & = q^{\nu (\sum_{j = 2}^l (m_{1j} + m_{j, \, l+1}) + 2 m_{1,l+1})} \, v_{\bm m}, \\
q^{\nu h_i} \, v_{\bm m} & = q^{\nu (\sum_{j = 1}^{i - 1} (m_{j i} - m_{j, \, i + 1}) - 2 m_{i, \, i + 1}  
- \sum_{j = i + 2}^{l + 1} (m_{i j} - m_{i + 1, \, j}))} v_{\bm m}, \\
e_0 \, v_{\bm m} & = q^{\sum_{i = 2}^l m_{i,l+1}} \, 
v_{{\bm m} + \epsilon_{1,l+1}}, \\
e_i \, v_{\bm m} & = - \kappa_q^{-1} \, 
q^{\sum_{j = i + 2}^{l + 1} (m_{i j} - m_{i + 1, \, j}) + m_{i,\, i + 1}} 
[m_{i, \, i + 1}]_q \, v_{{\bm m} - \epsilon_{i, i + 1}} \\
& \hspace{7em} - \sum_{j = i + 2}^{l + 1} 
q^{\sum_{k = j}^{l + 1} (m_{i k} - m_{i + 1, \, k}) - 1} \, [m_{i j}]_q \, 
v_{{\bm m} - \epsilon_{i j} + \epsilon_{i + 1, \, j}},
\end{align*}
where $i = 1, \ldots, l$. The $\uqbp$-module corresponding to $\rho''$ is denoted by $W''$. It is clear that the representation described by (\ref{5.4})--(\ref{5.7}) is $(\rho'')_{\widetilde \zeta}$. 

The representation $\rho''$ is reducible. Indeed, let us define an $l(l - 1)/2$-tuple $\bm p$ of nonnegative integers
\begin{equation}
\bm p = (p_{12}, \, p_{13}, \, p_{23}, \, \ldots, \, p_{1j}, \, \ldots, \, p_{j-1, \, j}, \, \ldots, \,
p_{1, \, l}, \, \ldots, \, p_{l - 1, \, l}), \label{bmp}
\end{equation}
such that
\begin{equation*}
\sum_{k = 1}^i (-1)^{i - k} p_{k j} \ge 0, \qquad 1 \le i < j \le l.
\end{equation*}
The subspaces generated by the vectors $v_{\bm m}$ with the indices restricted by the inequalities
\begin{equation*}
m_{i - 1, \, j} + m_{i j} \le p_{i j}, \qquad 1 \le i < j \le l,
\end{equation*}
where we assume that $m_{0 j} = 0$, are invariant with respect to the action of the quantum group $\uqbp$. We denote such $\uqbp$-submodule by $W''_{\bm p}$.

Now we introduce a partial order for the $l(l-1)/2$-tuples ${\bm p}$ of the form (\ref{bmp}) by assuming that $\bm p' < \bm p$ if $p'_{i j} < p_{i j}$ for all possible $i = 1, \ldots, l - 1$ and $j = 2, \ldots, l$.

Further, we denote by $\rho'$ the representation of $\uqbp$ defined by the
relations
\begin{align}
q^{\nu h_0} \, v_{\bm m} & = q^{\nu (2 m_{1} + \sum_{j = 2}^l m_j)} \, v_{\bm m}, 
\label{5.4b} \\
q^{\nu h_i} \, v_{\bm m} & = q^{\nu (m_{i + 1} - m_i)} v_{\bm m}, && i = 1, \ldots, l - 1, \label{5.5b} \\
q^{\nu h_l} \, v_{\bm m} & = q^{- \nu (2 m_l + \sum_{i = 1}^{l - 1} m_i)} \, v_{\bm m}, \label{5.6b} \\
e_0 \, v_{\bm m} & = q^{\sum_{j = 2}^l m_j} \,  v_{{\bm m} + \epsilon_{1}}, \label{5.7b} \\
e_i \, v_{\bm m} & = - q^{m_i - m_{i + 1} - 1} \, [m_i]_q \,
v_{{\bm m} - \epsilon_i + \epsilon_{i + 1}}, && i = 1, \ldots, l - 1, 
\label{5.8b} \\
e_l \, v_{\bm m} & = - \kappa_q^{-1} \, q^{m_{l}} \, [m_{l}]_q \,
v_{{\bm m} - \epsilon_{l}}, \label{5.9b}
\end{align}
where $\bm m$ denotes now the $l$-tuple of nonnegative integers $(m_1, \, \ldots, \, m_l)$, and ${\bm m} + \nu \epsilon_i$ means the respective shift of $m_i$ in this ordered tuple. The corresponding $\uqbp$-module is denoted by $W'$. We see that there are isomorphisms 
\begin{equation*}
W''_{\bm p} / \bigcup_{{\bm p'} \le {\bm p}} W''_{\bm p'} \cong W'[\xi_{\bm p}],
\end{equation*}
where the shift $\xi_{\bm p}$ is determined by the relations
\begin{align*}
\langle \xi_{\bm p}, \, h_0 \rangle & = \sum_{j = 2}^l p_{1 j}, \\
\langle \xi_{\bm p}, \, h_i \rangle & = \sum_{k = 1}^{i - 1} \sum_{j = 1}^k (-1)^{k - j}(p_{j i} - p_{j,\, i + 1}) - 2 \sum_{j = 1}^i (-1)^{i - j} p_{j, \, i + 1} \\
& \hspace{8em} {} - 2 \sum_{k = i + 2}^l \sum_{j = 1}^i (-1)^{i - j} p_{k j} + \sum_{k = i + 2}^l p_{i + 1, \, k}, 
\quad i = 1, \ldots, l - 1, \\
\langle \xi_{\bm p}, \, h_l \rangle & = \sum_{j = 1}^{l - 1} \sum_{k = 1}^j (-1)^{j - k} p_{k l}.
\end{align*}

It should be noted here that the integers $m_i$ in the new multi-index $\bm m$ in the module relations (\ref{5.4b})--(\ref{5.9b}) are nothing but the former $m_{i, \, l+1}$, $i = 1,\ldots,l$, survived the reduction to the factor module. We have denoted $m_{i, \, l+1}$ shortly by $m_i$ after the reduction,
and used a similar simplification also for the shift units $\epsilon_{i, \, l+1}$, for which we have reserved the notation $\epsilon_i$.

\section{\texorpdfstring{Interpretation in terms of $q$-oscillators}{Interpretation in terms of q-oscillators}} \label{s:9}

Degenerations of the shifted $\uqbp$-modules have a useful interpretation in terms of the so-called $q$-oscillators \cite{BazHibKho02, BooGoeKluNirRaz13, BooGoeKluNirRaz14a, NirRaz14, BooGoeKluNirRaz14b}. 
The $q$-oscillator algebra $\Osc_q$ is defined as a unital associative $\bbC$-algebra with generators $b^\dagger$, $b$, $q^{\nu N}$, $\nu \in \bbC$, and relations
\begin{gather*}
q^0 = 1, \qquad q^{\nu_1 N} q^{\nu_2 N} = q^{(\nu_1 + \nu_2)N}, \\
q^{\nu N} b^\dagger q^{-\nu N} = q^\nu b^\dagger, \qquad 
q^{\nu N} b q^{-\nu N} = q^{-\nu} b, \\
b^\dagger b = [N]_q, \qquad b b^\dagger = [N + 1]_q,
\end{gather*}
see, for example, section 5.1 of \cite{KliSch97} and references therein. Here, we again consider the deformation parameter to be $q = \exp\hbar$, where $\hbar$ is a complex number, such that $q$ is not a root of unity. 

There are two standard representations of the $q$-oscillator algebra. They 
are constructed as follows. One can see that the relations
\begin{gather*}
q^{\nu N} v_m = q^{\nu m} v_m, \\*
b^\dagger v_m = v_{m + 1}, \qquad b \, v_m = [m]_q v_{m - 1},
\end{gather*}
supplied with the assumption $v_{-1} = 0$, endow the free vector space generated by the set $\{ v_0, \, v_1, \, \ldots \, \}$ with the structure of an $\Osc_q$-module. We denote this $\Osc_q$-module by $W^+$ and the corresponding representation by $\chi^+$. The other representation of $\Osc_q$ is defined by the relations
\begin{gather*}
q^{\nu N} v_m = q^{-\nu (m + 1)} v_m,  \\
b \, v_m = v_{m + 1}, \qquad b^\dagger \, v_m = - [m]_q v_{m - 1},
\end{gather*}
where it is assumed again that $v_{-1} = 0$. Similarly as before, these relations endow the free vector space generated by the set  $\{ v_0, \, v_1, \, \ldots \, \}$ with the structure of an $\Osc_q$-module. This $\Osc_q$-module and the corresponding representation are denoted by $W^-$ and $\chi^-$, respectively. However, since the automorphism of $\Osc_q$ 
\begin{equation*}
b \to b^\dagger, \qquad b^\dagger \to - b, \qquad q^{\nu N} \to q^{-\nu(N + 1)} 
\end{equation*}
relates these representations, it is actually sufficient to use only one of them.

We consider the tensor product of $l$ copies of the $q$-oscillator algebra, 
$\Osc_q \otimes \ldots \otimes \Osc_q = (\Osc_q)^{\otimes \, l}$, and denote
\begin{gather*}
b_i = 1 \otimes \ldots \otimes b \otimes \ldots \otimes 1, \qquad
b^\dagger_i = 1 \otimes \ldots \otimes b^\dagger \otimes \ldots \otimes 1, \\
q^{\nu N_i} = 1 \otimes \ldots \otimes q^{\nu N} \otimes \ldots \otimes 1, 
\end{gather*}
where $b$, $b^\dagger$ and $q^{\nu N}$ occupy only the $i$-th place of
the respective tensor products.

Let us consider the $\uqbp$-module $W'$ and the corresponding representation $\rho'$ given by relations (\ref{5.4b})--(\ref{5.9b}). Supply $W'$ with the structure of $(\Osc_q)^{\otimes \, l}$-module assuming 
that
\begin{gather*}
q^{\nu N_i} v_{\bm m} = q^{\nu m_i} v_{\bm m},  \\
b^\dagger_i \, v_{\bm m} = v_{\bm m + \epsilon_i}, \qquad 
b^{}_i \, v_{\bm m} = [m_i]_q v_{\bm m - \epsilon_i}.
\end{gather*}
Now, the module relations (\ref{5.4b})--(\ref{5.9b}) can be written in terms of the $q$-oscillators as 
follows:
\begin{align*}
q^{\nu h_0} \, v_{\bm m} & = q^{\nu (2 N_1 + \sum_{j = 2}^l N_j)} \, v_{\bm m}, \\
q^{\nu h_i} \, v_{\bm m} & = q^{\nu (N_{i + 1} - N_{i})} \, v_{\bm m}, && i = 1, \ldots, l-1, \\
q^{\nu h_l} \, v_{\bm m} & = q^{- \nu (2 N_l + \sum_{j = 1}^{l - 1} N_j)} \, v_{\bm m}, \\
e_0 \, v_{\bm m} & = b^\dagger_1 \, q^{\sum_{j = 2}^l N_j} \, v_{{\bm m}}, \\
e_i \, v_{\bm m} & = - b^\ms_i \, b^{\mathstrut \dagger}_{i + 1} \, q^{N_i - N_{i + 1} - 1} \, v_{{\bm m}}, && i = 1, \ldots, l - 1, \\
e_l \, v_{\bm m} & = - \kappa_q^{-1} \, b^\ms_l \, q^{N_l} \, v_{\bm m}.
\end{align*} 
It is natural now to define a homomorphism $\rho : \uqbp \to \Osc_q^{\otimes \, l}$ by the relations
\begin{align*}
& \rho(q^{\nu h_0}) = q^{\nu (2 N_1 + \sum_{j = 2}^l N_j)}, &&
\rho(q^{\nu h_i}) = q^{\nu (N_{i + 1} - N_{i})}, &&
\rho(q^{\nu h_l}) = q^{- \nu (2 N_l + \sum_{j = 1}^{l - 1} N_j)}, \\
& \rho(e_0) = b^\dagger_1 \, q^{\sum_{j = 2}^l N_j}, &&
\rho(e_i) = - b^\ms_i \, b^{\mathstrut \dagger}_{i + 1} \, q^{N_i - N_{i + 1} - 1}, &&
\rho(e_l) = - \kappa_q^{-1} \, b^\ms_l \, q^{N_l},
\end{align*}
where $i = 1, \ldots, l - 1$. In a sense, the homomorphism $\rho$ plays the role of the Jimbo's homomorphism $\varepsilon$. To get a representation of $\uqbp$, one chooses some representation of $(\Osc_q)^{\otimes \, l}$, and then takes the composition of this representation with $\rho$. In particular, for the representation $\rho'$ one has
\begin{equation*}
\rho' = (\chi^+ \otimes \cdots \otimes \chi^+) \circ \rho \circ \Gamma_\zeta.
\end{equation*}
More representations can be obtained using twisting by the automorphisms of $\uqbp$. These are the representations used for the construction of the $Q$-operators.

\section{Conclusions} \label{s:10}

We have analysed the Verma modules over the quantum group $\uqgllpo$ for arbitrary values of $l$. The explicit expressions for the action of the generators on the elements of the natural basis have been obtained. The corresponding representations of the quantum loop algebras $\uqlsllpo$ have been constructed. This has allowed us to find certain representations of the positive Borel subalgebras of $\uqlsllpo$ as degenerations of the shifted representations. These are the representations used in the construction of the so-called $Q$-operators in the theory of quantum integrable systems. The interpretation of the corresponding simple quotient modules in terms of representations of the $q$-deformed oscillator algebra has been given. The obtained results can be used for the investigation of quantum integrable systems in the spirit of the papers \cite{BazLukZam96, BazLukZam97, BazLukZam99, BazHibKho02} and \cite{BooGoeKluNirRaz13, BooGoeKluNirRaz14a, NirRaz14, BooGoeKluNirRaz14b}. We expect also applications to the higher rank generalization of the quantum group approach to the construction of correlation functions for integrable spin chain and conformally invariant models \cite{BooJimMiwSmiTak07, BooJimMiwSmiTak09, JimMiwSmi09, BooJimMiwSmi10, JimMiwSmi11}.

The $q$-oscillator representations of the Borel subalgebras of $\uqlsllpo$ are closely related to the prefundamental representations introduced by D.~Hernandez and M.~Jimbo \cite{HerJim12}. The explicit relation for the case of $\uqlslii$ and $\uqlsliii$ was found in the paper \cite{BooGoeKluNirRaz15}.

\vskip 1.5em

{\em Acknowledgements.\/} The authors are grateful to H.~Boos, F.~G\"ohmann and A.~Kl\"umper for discussions. This work was supported in part by the DFG grant KL \hbox{645/10-1} and by 
the Volkswagen Foundation. It was also supported in part by the RFBR grants \#~14-01-91335 
and \#~16-01-00473.

\appendix

\section*{\texorpdfstring{Appendix. Acting by $E_{1, \, l + 1}$ on the basis vectors}{Appendix. Acting by E1,l + 1 on the basis vectors}} \label{a:1}

\stepcounter{section}

Here we derive the action of the root vector $E_{1, \, l + 1}$ on the basis of the Verma $U_q(\gllpo)$-module defined in section \ref{s:4}. To this end, we first obtain from equations (\ref{2.9})--(\ref{2.16}) and (\ref{2.18})--(\ref{2.25}) the following subsidiary formulas:
\begin{align*}
& E_{i, \, l + 1}^\ms \, F_{i j}^{m_{i j}} = F_{i j}^{m_{i j}} \, E_{i, \, l + 1}^\ms - q^{- m_{i j} + 1} \, [m_{i j}^\ms]_q \, F_{i j}^{m_{i j} - 1} \, E_{j, \, l + 1}^\ms \, q^{H_{i j}}, \quad 1 \le i < j \le l, \\
& E_{i, \, l + 1}^\ms \, F_{i, \, l + 1}^{m_{i, \, l + 1}} = F_{i,\, l + 1}^{m_{i, \, l + 1}} \, 
E_{i, \, l + 1}^\ms + [m_{i, \, l + 1}^\ms]_q \, F_{i, \, l + 1}^{m_{i, \, l + 1} - 1} \, [H_{i, \, l + 1}^\ms - m_{i, \, l + 1}^\ms + 1]_q, \quad 1 \le i \le l, \\
& E_{j, \, l + 1}^\ms \, F_{i k}^{m_{i k}} = F_{i k}^{m_{i k}} \, E_{j, \, l + 1}^\ms - \kappa_q \, [m_{i k}^\ms]_q \, F_{i j}^\ms \, F_{i k}^{m_{i k} - 1} \, E_{k, \, l + 1}^\ms \, q^{H_{j k}}, \quad 1 \le i < j < k \le l, \\
& E_{j, \, l + 1}^\ms \, F_{i, \, l + 1}^{m_{i, \, l + 1}} = F_{i, \, l + 1}^{m_{i, \, l + 1}} \, E_{j, \, l + 1}^\ms + [m_{i, \, l + 1}^\ms]_q \, F_{i j}^\ms \, F_{i, \, l + 1}^{m_{i, \, l + 1} - 1} \, q^{H_{j, \, l + 1}}, \quad 1 \le i < j \le l, \\
& E_{i, \, l + 1}^\ms \, F_{j, \, l + 1}^{m_{j, \, l + 1}} = F_{j, \, l + 1}^{m_{j, \, l + 1}} \, E_{i, \, l + 1}^\ms + q^{m_{j, \, l + 1}} \, [m_{j, \, l + 1}^\ms]_q \, F_{j, \, l + 1}^{m_{j, \, l + 1} - 1} \, E_{i j}^\ms \, q^{- H_{j, \, l + 1}}, \quad 1 \le i < j \le l,
\end{align*}
In the first equation of these five we take into account that $E_{j, \, l + 1} \, F_{i j} = F_{i j} \, E_{j, \, l + 1}$ and
\begin{equation*}
q^{H_{i j}} \, F_{i j} = q^{-2} \, F_{i j} \, q^{H_{i j}}, \qquad E_{j, \, l + 1} \, q^{H_{i j}} 
= q^{-2} \, q^{H_{i j}} \, E_{j, \, l + 1}.
\end{equation*}
These commutations are needed to move the root vectors $E_{i, \, l + 1}$, $1 \le i \le l$, 
to the right, where they annihilate the highest weight vector $v_{\bm 0}$. Due to the last relation a root vector $E_{i j}$ with $1 \le i < j \le l$ appears. However, it commutes 
with the remaining root vectors $F_{i j}$ and annihilates $v_{\bm 0}$. 

Besides, one needs relations which enable the root vectors $F_{i j}$ arising during the commutations to move back to the left to occupy a proper position according to the prescribed ordering. Such relations are
\begin{align*}
& F_{j k}^{m_{j k}} \, F_{i j}^\ms = q^{-m_{j k}} \, F_{i j}^\ms \, F_{j k}^{m_{j k}} + [m_{j k}^\ms]_q \, F_{i k}^\ms \, F_{j k}^{m_{j k}-1}, \quad 1 \le i < j < k \le l + 1, \\
& F_{j n}^{m_{j n}} \, F_{i k}^\ms = F_{i k}^\ms \, F_{j n}^{m_{j n}} + \kappa_q \, q^{m_{j n} - 1} \, [m_{j n}^\ms]_q \, F_{i n}^\ms \, F_{j k}^\ms \, F_{j n}^{m_{j n} - 1}, \quad 1 \le i < j < k < n \le l + 1.
\end{align*}

To describe the final result, let us introduce some additional notations. First of all for $0 \le k \le l - 1$ denote
\begin{equation*}
\Lambda_{l, \, k} = \{ (i_0 = 1, \, i_1, \, \ldots, \, i_k, \, i_{k+1} = l + 1) \in \bbN^{\times (k + 2)} \mid i_0 < i_1 < \ldots < i_k < i_{k + 1} \}.
\end{equation*}
Note that the first and last elements of the tuples entering $\Lambda_{l, \, k}$ are fixed. However, it is convenient to consider the tuples of this form. Up to this extension, the set $\Lambda_l$ introduced in section \ref{s:3} coincides with $\Lambda_{l, \, 2}$. To make the formulas more compact we denote an element $(i_0, \, i_1, \, \ldots, \, i_k, \, i_{k + 1})$ of $\Lambda_{l, \, k}$ as $\bm i$. Further, we use the notation
\begin{equation*}
\Psi_{l, \, k} = \{ (\bm i, \, \bm j) \in \Lambda_{l, \, k} \times \Lambda_{l, \, k} \mid j_{a - 1} < i_a \le j_a, \ a = 1, \ldots k \}.
\end{equation*}

Now we can write
\begin{multline}
E_{1, \, l + 1} v_{\bm m} = \sum_{k = 0}^{l - 1} \sum_{\substack{(\bm i, \, \bm j) \in \Psi_{l, \, k} \\ i_k = j_k}} a_{\bm m | k, \, \bm i, \, \bm j} \, v_{\bm m - \epsilon_{j_0 j_1} + \epsilon_{i_1 j_1}  - \ldots - \epsilon_{j_{k - 1} j_k} + \epsilon_{i_k j_k} - \epsilon_{j_k j_{k + 1}}}^\ms \\
+ \sum_{k = 1}^{l - 1} \sum_{\substack{(\bm i, \, \bm j) \in \Psi_{l, \,k} \\ i_k \ne j_k}} b_{\bm m | k, \, \bm i, \, \bm j} \, v_{\bm m - \epsilon_{j_0 j_1} + \epsilon_{i_1 j_1}  - \ldots - \epsilon_{j_{k - 1} j_k} + \epsilon_{i_k j_k} - \epsilon_{j_k j_{k + 1}}}^\ms. \label{E1l1}
\end{multline}
Here we assume that $\epsilon_{i i} = 0$. The explicit form of the coefficients $a_{\bm m | k, \, \bm i, \, \bm j}$ and $b_{\bm m | k, \, \bm i, \, \bm j}$ is
\begin{align*}
& a_{\bm m | k, \, \bm i, \, \bm j} = (- 1)^k  \kappa_q^{\gamma_{\bm i, \, \bm j}} q^{\lambda_1 - \lambda_{j_k} - \sum_{j = 2}^{l + 1} m_{1 j} + m_{j_k, \, l + 1} - \sum_{a = 1}^k \delta_{\bm m | a, \, \bm i, \, \bm j} + k - \gamma_{\bm i, \, \bm j}} \\
& \hspace{10em} {} \times [\lambda_{j_k} - \lambda_{l + 1} - \sum_{j = j_k + 1}^l m_{i_k j} - \sum_{i = i_k + 1}^l m_{i, \, l + 1} + 1]_q \prod_{a = 1}^{k + 1} [m_{i_{a -1} j_a}]_q, \\
& b_{\bm m | k, \, \bm i, \, \bm j} = (- 1)^k  \kappa_q^{\gamma_{\bm i, \, \bm j} - 1} q^{\lambda_1 - \lambda_{l + 1} - \sum_{j = 2}^{l + 1} m_{1 j} - \sum_{a = 1}^k \delta_{\bm m | a, \, \bm i, \, \bm j} + k - \gamma_{\bm i, \, \bm j}} \prod_{a = 1}^{k + 1} [m_{i_{a -1} j_a}]_q,
\end{align*}
where we denote
\begin{equation*}
\gamma_{\bm i, \, \bm j} = \sharp (\{ a = 1, \ldots, k \mid i_a \ne j_a \})
\end{equation*}
and
\begin{equation*}
\delta_{\bm m | a, \, \bm i, \, \bm j} = \sum_{j = j_{a - 1} + 1}^{j_a - 1} m_{i_{a - 1} j} + \sum_{i = i_{a - 1} + 1}^{i_a} m_{i j_a}.
\end{equation*}
It is also assumed that $m_{i i} = 0$.

\providecommand{\href}[2]{#2}


\begin{thebibliography}{10}

\bibitem{Dri85}
V.~G. Drinfeld, \emph{Hopf algebras and quantum {Y}ang--{B}axter equation},
  Soviet Math. Dokl. \textbf{32} (1985), 254--258.

\bibitem{Dri87}
V.~G. Drinfeld, \emph{Quantum groups}, Proceedings of the International
  Congress of Mathematicians, Berkeley, 1986 (A.~E. Gleason, ed.), vol.~1,
  American Mathematical Society, Providence, 1987, pp.~798--820.

\bibitem{Jim85}
M.~Jimbo, \emph{A $q$-difference analogue of {$\mathrm U(\mathfrak g)$} and the
  {Y}ang-{B}axter equation}, \href{http://dx.doi.org/10.1007/BF00704588}{Lett.
  Math. Phys.} \textbf{10} (1985), 63--69.

\bibitem{Bax72a}
R.~J. Baxter, \emph{{P}artition function of the {E}ight-{V}ertex
  lattice model}, \href{http://dx.doi.org/10.1016/0003-4916(72)90335-1}{Ann.
  Phys. (N. Y.)} \textbf{70} (1972), 193--228.

\bibitem{KluPea92}
A.~Kl\"umper and P.~A. Pearce, \emph{Conformal weights of {RSOS} lattice models
  and their fusion hierarchies},
  \href{http://dx.doi.org/10.1016/0378-4371(92)90149-K}{Physica A} \textbf{183}
  (1992), 304--350.

\bibitem{KunNakSuz94}
A.~Kuniba, T.~Nakanishi, and J.~Suzuki, \emph{Functional relations in solvable
  lattice models. {I}. {F}unctional relations and representation theory},
  \href{http://dx.doi.org/10.1142/S0217751X94002119}{Int. J. Mod. Phys. A}
  \textbf{9} (1994), 5215--5266,
  \href{http://arxiv.org/abs/hep-th/9309137}{{\tt arXiv:hep-th/9309137}}.

\bibitem{Koj08}
T.~Kojima, \emph{Baxter's ${Q}$-operator for the ${W}$-algebra ${W_N}$},
  \href{http://dx.doi.org/10.1088/1751-8113/41/35/355206}{J. Phys. A: Math.
  Theor} \textbf{41} (2008), 355206 (16pp),
  \href{http://arxiv.org/abs/0803.3505}{{\tt arXiv:0803.3505 [nlin.SI]}}.

\bibitem{BazLukMenSta10}
V.~V. Bazhanov, T.~{\L}ukowski, C.~Meneghelli, and M.~Staudacher, \emph{A
  shortcut to the {Q}-operator},
  \href{http://dx.doi.org/10.1088/1742-5468/2010/11/P11002}{J. Stat. Mech.}
  (2010), P11002 (37pp), \href{http://arxiv.org/abs/1005.3261}{{\tt
  arXiv:1005.3261 [hep-th]}}.

\bibitem{Tsu10}
Z.~Tsuboi, \emph{Solutions of the {$T$}-system and {B}axter equations for
  supersymmetric spin chains},
  \href{http://dx.doi.org/10.1016/j.nuclphysb.2009.08.009}{Nucl. Phys. B}
  \textbf{626} (2010), 399--455, \href{http://arxiv.org/abs/0906.2039}{{\tt
  arXiv:0906.2039 [math-ph]}}.

\bibitem{FraLukMenSta11}
R.~Frassek, T.~{\L}ukowski, C.~Meneghelli, and M.~Staudacher, \emph{Oscillator
  construction of $\mathfrak{su}(n|m)$ $q$-operators},
  \href{http://dx.doi.org/10.1016/j.nuclphysb.2011.04.008}{Nucl. Phys. B}
  \textbf{850} (2011), 175--198, \href{http://arxiv.org/abs/1012.6021}{{\tt
  arXiv:1012.6021 [math-ph]}}.

\bibitem{KazLeuTsu12}
V.~Kazakov, S.~Leurent, and Z.Tsuboi, \emph{Baxter's ${Q}$-operators and
  operatorial {B}\"acklund flow for quantum (super)-spin chains},
  \href{http://dx.doi.org/10.1007/s00220-012-1428-9}{Commun. Math. Phys.}
  \textbf{311} (2012), 787--814, \href{http://arxiv.org/abs/1010.4022}{{\tt
  arXiv:1010.4022 [math-ph]}}.

\bibitem{AleKazLeuTsuZab13}
A.~Alexandrov, V.~Kazakov, S.~Leurent, Z.~Tsuboi, and A.~Zabrodin,
  \emph{Classical tau-function for quantum spin chains},
  \href{http://dx.doi.org/10.1007/JHEP09(2013)064}{JHEP} \textbf{09} (2013),
  064 (66pp), \href{http://arxiv.org/abs/1112.3310}{{\tt arXiv:1112.3310
  [math-ph]}}.

\bibitem{Tsu13a}
Z.~Tsuboi, \emph{Wronskian solutions of the ${T}$-, ${Q}$- and ${Y}$-systems
  related to infinite dimensional unitarizable modules of the general linear
  superalgebra $gl({M}|{N})$},
  \href{http://dx.doi.org/10.1016/j.nuclphysb.2013.01.007}{Nucl. Phys. B}
  \textbf{870} (2013), 92--137, \href{http://arxiv.org/abs/1109.5524}{{\tt
  arXiv:1109.5524 [hep-th]}}.

\bibitem{FraLukMenSta13}
R.~Frassek, T.~{\L}ukowski, C.~Meneghelli, and M.~Staudacher, \emph{Baxter
  operators and {H}amiltonians for ``nearly all'' integrable closed
  $\mathfrak{gl}(n)$ spin chains},
  \href{http://dx.doi.org/10.1016/j.nuclphysb.2013.06.006}{Nucl. Phys. B}
  \textbf{874} (2013), 620--646, \href{http://arxiv.org/abs/1112.3600}{{\tt
  arXiv:1112.3600 [math-ph]}}.

\bibitem{BazLukZam96}
V.~V. Bazhanov, S.~L. Lukyanov, and A.~B. Zamolodchikov, \emph{Integrable
  structure of conformal field theory, quantum {K}d{V} theory and thermodynamic
  {B}ethe ansatz}, \href{http://dx.doi.org/10.1007/BF02101898}{Commun. Math.
  Phys.} \textbf{177} (1996), 381--398,
  \href{http://arxiv.org/abs/hep-th/9412229}{{\tt arXiv:hep-th/9412229}}.

\bibitem{BazLukZam97}
V.~V. Bazhanov, S.~L. Lukyanov, and A.~B. Zamolodchikov, \emph{Integrable
  structure of conformal field theory {II}. {Q}-operator and {DDV} equation},
  \href{http://dx.doi.org/10.1007/s002200050240}{Commun. Math. Phys.}
  \textbf{190} (1997), 247--278,
  \href{http://arxiv.org/abs/hep-th/9604044}{{\tt arXiv:hep-th/9604044}}.

\bibitem{BazLukZam99}
V.~V. Bazhanov, S.~L. Lukyanov, and A.~B. Zamolodchikov, \emph{Integrable
  structure of conformal field theory {III}. {T}he {Y}ang--{B}axter relation},
  \href{http://dx.doi.org/10.1007/s002200050531}{Commun. Math. Phys.}
  \textbf{200} (1999), 297--324,
  \href{http://arxiv.org/abs/hep-th/9805008}{{\tt arXiv:hep-th/9805008}}.

\bibitem{BazHibKho02}
V.~V. Bazhanov, A.~N. Hibberd, and S.~M. Khoroshkin, \emph{Integrable structure
  of {$\mathcal W_3$} conformal field theory, quantum {B}oussinesq theory and
  boundary affine {T}oda theory},
  \href{http://dx.doi.org/10.1016/S0550-3213(01)00595-8}{Nucl. Phys. B}
  \textbf{622} (2002), 475--574,
  \href{http://arxiv.org/abs/hep-th/0105177}{{\tt arXiv:hep-th/0105177}}.

\bibitem{BooGoeKluNirRaz13}
H.~Boos, F.~G{\"o}hmann, A.~Kl{\"u}mper, Kh.~S. Nirov, and A.~V. Razumov,
  \emph{Universal integrability objects},
  \href{http://dx.doi.org/10.1007/s11232-013-0002-8}{Theor. Math. Phys.}
  \textbf{174} (2013), 21--39, \href{http://arxiv.org/abs/1205.4399}{{\tt
  arXiv:1205.4399 [math-ph]}}.

\bibitem{BooGoeKluNirRaz14a}
H.~Boos, F.~G{\"o}hmann, A.~Kl\"umper, Kh.~S. Nirov, and A.~V. Razumov,
  \emph{Universal ${R}$-matrix and functional relations},
  \href{http://dx.doi.org/10.1142/S0129055X14300052}{Rev. Math. Phys.}
  \textbf{26} (2014), 1430005 (66pp),
  \href{http://arxiv.org/abs/1205.1631}{{\tt arXiv:1205.1631 [math-ph]}}.

\bibitem{NirRaz14}
Kh.~S. Nirov and A.~V. Razumov, \emph{Quantum groups and functional relations
  for lower rank}, \href{http://dx.doi.org/10.1016/j.geomphys.2016.10.014}{J.
  Geom. Phys.} \textbf{112} (2017), 1--28,
  \href{http://arxiv.org/abs/1412.7342}{{\tt arXiv:1412.7342 [math-ph]}}.

\bibitem{BooGoeKluNirRaz14b}
H.~Boos, F.~G{\"o}hmann, A.~Kl\"umper, Kh.~S. Nirov, and A.~V. Razumov,
  \emph{Quantum groups and functional relations for higher rank},
  \href{http://dx.doi.org/10.1088/1751-8113/47/27/275201}{J. Phys. A: Math.
  Theor.} \textbf{47} (2014), 275201 (47pp),
  \href{http://arxiv.org/abs/1312.2484}{{\tt arXiv:1312.2484 [math-ph]}}.

\bibitem{Jim86a}
M.~Jimbo, \emph{A $q$-analogue of {$\mathrm U(\mathfrak{gl}(N + 1))$}, {H}ecke
  algebra, and the {Y}ang--{B}axter equation},
  \href{http://dx.doi.org/10.1007/BF00400222}{Lett. Math. Phys.} \textbf{11}
  (1986), 247--252.

\bibitem{LezSav74}
A.~N. Leznov and M.~V. Saveliev, \emph{A parametrization of compact groups},
  \href{http://dx.doi.org/10.1007/BF01075497}{Funt. Anal. Appl.} \textbf{8}
  (1974), 347--348.

\bibitem{AshSmiTol79}
R.~M. Asherova, Yu.~F. Smirnov, and V.~N. Tolstoy, \emph{Description of a class
  of projection operators for semisimple complex lie algebras},
  \href{http://dx.doi.org/10.1007/BF01140268}{Math. Notes} \textbf{26} (1979),
  499--504.

\bibitem{Tol89}
V.~N. Tolstoy, \emph{Extremal projections for contragredient {L}ie algebras and
  superalgebras of finite growth},
  \href{http://dx.doi.org/10.1070/RM1989v044n01ABEH002023}{Russian Math.
  Surveys} \textbf{44} (1989), 267--258.

\bibitem{Yam89}
H.~Yamane, \emph{A {P}oincar\'e--{B}irkhoff--{W}itt theorem for quantized
  universal enveloping algebras of type {$A_N$}},
  \href{http://dx.doi.org/10.2977/prims/1195173355}{Publ. RIMS. Kyoto Univ.}
  \textbf{25} (1989), 503--520.

\bibitem{BooGoeKluNirRaz15}
H.~Boos, F.~G{\"o}hmann, A.~Kl\"umper, Kh.~S. Nirov, and A.~V. Razumov,
  \emph{Oscillator versus prefundamental representations},
  \href{http://dx.doi.org/10.1063/1.4966925}{J. Math. Phys.} \textbf{57}
  (2016), 111702 (23pp), \href{http://arxiv.org/abs/1512.04446}{{\tt
  arXiv:1512.04446 [math-ph]}}.
  
\bibitem{Kac90}
V.~Kac, \emph{Infinite dimensional {L}ie algebras}, Cambridge University Press,
  Cambridge, 1990.

\bibitem{ChaPre91}
V.~Chari and A.~Pressley, \emph{Quantum affine algebras},
  \href{http://dx.doi.org/10.1007/BF02102063}{Commun. Math. Phys.} \textbf{142}
  (1991), 261--283.

\bibitem{ChaPre94}
V.~Chari and A.~Pressley, \emph{A guide to quantum groups}, Cambridge
  University Press, Cambridge, 1994.

\bibitem{KliSch97}
A.~Klimyk and K.~Schm{\"u}dgen, \emph{Quantum groups and their
  representations}, Texts and Monographs in Physics, Springer, Heidelberg,
  1997.

\bibitem{BooJimMiwSmiTak07}
H.~Boos, M.~Jimbo, T.~Miwa, F.~Smirnov, and Y.~Takeyama, \emph{Hidden
  {G}rassmann structure in the {XXZ} model},
  \href{http://dx.doi.org/10.1007/s00220-007-0202-x}{Commun. Math. Phys.}
  \textbf{272} (2007), 263--281,
  \href{http://arxiv.org/abs/hep-th/0606280}{{\tt arXiv:hep-th/0606280}}.

\bibitem{BooJimMiwSmiTak09}
H.~Boos, M.~Jimbo, T.~Miwa, F.~Smirnov, and Y.~Takeyama, \emph{Hidden
  {G}rassmann structure in the {XXZ} model {II}: {C}reation operators},
  \href{http://dx.doi.org/10.1007/s00220-008-0617-z}{Commun. Math. Phys.}
  \textbf{286} (2009), 875--932, \href{http://arxiv.org/abs/0801.1176}{{\tt
  arXiv:0801.1176 [hep-th]}}.

\bibitem{JimMiwSmi09}
M.~Jimbo, T.Miwa, and F.~Smirnov, \emph{Hidden {G}rassmann structure in the
  {XXZ} model {III}: {I}ntroducing the {M}atsubara direction},
  \href{http://dx.doi.org/10.1088/1751-8113/42/30/304018}{J. Phys. A: Math.
  Theor.} \textbf{42} (2009), 304018 (31pp),
  \href{http://arxiv.org/abs/0811.0439}{{\tt arXiv:0811.0439 [math-ph]}}.

\bibitem{BooJimMiwSmi10}
H.~Boos, M.~Jimbo, T.~Miwa, and F.~Smirnov, \emph{Hidden {G}rassmann structure
  in the {XXZ} model {IV}: {CFT} limit},
  \href{http://dx.doi.org/10.1007/s00220-010-1051-6}{Commun. Math. Phys.}
  \textbf{299} (2010), 825--866, \href{http://arxiv.org/abs/0911.3731}{{\tt
  arXiv:0911.3731 [hep-th]}}.

\bibitem{JimMiwSmi11}
M.~Jimbo, T.~Miwa, and F.~Smirnov, \emph{Hidden {G}rassmann structure in the
  {XXZ} model {V}: sine-{G}ordon model},
  \href{http://dx.doi.org/10.1007/s11005-010-0438-9}{Lett. Math. Phys.}
  \textbf{96} (2011), 325--365, \href{http://arxiv.org/abs/1007.0556}{{\tt
  arXiv:1007.0556 [hep-th]}}.

\bibitem{HerJim12}
D.~Hernandez and M.~Jimbo, \emph{Asymptotic representations and {D}rinfeld
  rational fractions}, \href{http://dx.doi.org/10.1112/S0010437X12000267}{Comp.
  Math.} \textbf{148} (2012), 1593--1623,
  \href{http://arxiv.org/abs/1104.1891}{{\tt arXiv:1104.1891 [math.QA]}}.

\end{thebibliography}
\end{document}